\documentclass[12pt,aps,floats,showpacs,amssymb,tightenlines]{revtex4}

\usepackage{amsmath}
\usepackage{amsfonts}
\usepackage{amscd}
\usepackage{epsfig}
\usepackage{amssymb}
\usepackage{tabularx}
\def\a{\alpha}

\def\g{\gamma}

\def\m{\mu}

\def\d{\delta}
\def\l{\lambda}
\def\L{\Lambda}

\def\w{\omega}
\def\c{\nabla}
\def\p{\partial}
\def\k{\kappa}

\begin{document}

\title{Linear stability of Einstein-Gauss-Bonnet static spacetimes  \\
Part I: tensor perturbations}
\author{Gustavo Dotti and Reinaldo J. Gleiser}
\affiliation{Facultad de Matem\'atica, Astronom\'{\i}a y F\'{\i}sica,
Universidad Nacional de C\'ordoba, Ciudad Universitaria,
(5000) C\'ordoba, Argentina}
\email{gdotti@fis.uncor.edu}

\begin{abstract}
 We study the stability under linear perturbations
 of a class of static solutions of 
 Einstein-Gauss-Bonnet 
gravity in $D=n+2$ dimensions with spatial slices of the form 
$\Sigma_{\k}^n \, \times \, {\mathbb R}^+$, $\Sigma_{\k}^n$ 
an $n-$manifold of constant curvature $\k$. Linear perturbations 
for this class of space-times can be 
generally classified into tensor, vector and scalar 
types. The analysis in this paper is restricted to tensor perturbations.
We show that the evolution equations for tensor perturbations can be cast in
Schr\"odinger form, and obtain the exact potential. We use 
$S-$deformations to analyze 
the hamiltonian spectrum, and find an S-deformed  potential that factors 
in a convenient way, allowing us to draw  definite 
conclusions about stability in every case. It is found that there is 
a minimal mass for a $D=6$ black hole with a positive curvature horizon 
to be stable. For any $D$, there is also a critical mass {\em above} which 
black holes with  negative curvature horizons are unstable.
\end{abstract}

\maketitle

\section{Introduction} \label{intro}
The analysis of the properties and behavior of gravity 
in higher dimensions has become in recent years a major area of research, 
motivated in particular by developments in string theory. Among others,
the Einstein-Gauss-Bonnet (EGB) gravity theory has been  
singled out as relevant to the low energy string limit \cite{str1}.
The EGB lagrangian is a linear combination of Euler densities continued
from lower dimensions. It gives equations involving  up to 
second order derivatives of the metric,  and has the  same degrees
of freedom as ordinary Einstein theory. An appropriate 
 choice of  the coefficients 
in front of the  Euler densities enlarges 
the local Lorentz symmetry to  local $(A)dS$
symmetry \cite{ads1,ads2}.
A number of  solutions to the EGB equations, many of
them relevant to the development of the $AdS-CFT$ correspondence \cite{jm}, are
known, among them 
a variety of black holes in asymptotically Euclidean or $(A)dS$ spacetimes
\cite{w1,w2,w3,atz,chm}.
 These
were found mostly because they are highly  symmetric.
Analyzing their linear stability, however, confronts us with the  complexity of
the EGB  equations, since the perturbative terms break the simplifying symmetries of
the background metric. The linear stability under tensor perturbations 
of higher dimensional static black holes 
in Einstein gravity was studied in \cite{gh}; the stability of higher dimensional 
rotating Einstein black holes is analyzed in \cite{cardoso}.
The quasinormal modes of 
higher dimensional 
black holes are analyzed in \cite{berti} for Einstein gravity 
and in \cite{gm} for EGB gravity.\\
In this paper we consider spacetimes that admit locally a metric of the form 
\begin{equation} \label{b}
ds^2 = -f(r) dt^2 + g(r) dr^2 + r^2 \bar g _{ij} dx^i dx^j, 
\end{equation}
where $\bar g _{ij} dx^i dx^j$  is the line element of an $n-$dimensional 
 manifold $\Sigma_{\k}^n$ of constant curvature $\k=1,0$ or $-1$. 
Linear perturbations around (\ref{b}) can be conveniently classified,
following the scheme proposed in \cite{koda}, into
tensor, vector, and scalar perturbations.
The $\k=1$ case $\Sigma_{1}^n = S^n$ gives, for appropriate $f$ and $g$, 
cosmological solutions, as well as higher dimensional 
Schwarzchild black holes.
The stability of these solutions under tensor perturbations
was studied in \cite{dg}. In this paper we provide 
the details of the  calculations leading to the results in \cite{dg} 
as we extend them to the cases $\k=-1,0$. In Section \ref{tp} and 
Appendix A we 
introduce tensor perturbations  around (\ref{b}) and calculate 
the variation of the Riemann tensor, then in Section \ref{egb} 
we review the basics of Einstein-Gauss-Bonnet theory (EGB), 
exhibit known solutions
 of the form (\ref{b}) from \cite{w1,w2,w3,atz}, 
obtain the perturbative equation for the tensor mode, and 
reduce it to a  Schr\"odinger 
equation.
In Section \ref{cms} we classify the EGB solutions (\ref{b}).
 A number of different possibilities arise depending on the spacetime 
dimension, the value of the cosmological constant and the strength of the 
coupling of the Gauss-Bonnet term (string-tension). 
Compact manifolds $\Sigma$
of negative (null) curvature can be obtained by taking quotients 
of hyperbolic space (Euclidean space) by appropriate discrete isometry groups, 
and black holes having such manifolds as horizons can be constructed in 
EGB   gravity (for black holes with exotic horizons see, e.g., \cite{atz}). The stability of cosmologies and 
black hole 
solutions  is studied in Section \ref{sms} 
using the {\em S-deformation} approach \cite{ki}. 
In spite of the complexity of the original Schr\"odinger 
potential, an S-deformed  potential is found that factors 
in a convenient way and allows us to draw  definite 
conclusions about stability in every case. 
Our preliminary work on vector and scalar perturbations
\cite{dg3} seems to indicate that this factorization is peculiar of 
the tensor mode.  Conclusions about tensor perturbations 
can be found in Section \ref{conc}.

\section{Tensor perturbations of a class of static spacetimes} \label{tp}
\noindent
As stated in the previous Section, in this paper we consider spacetimes with 
metrics locally given by (\ref{b}). We use $a,b,c,d,...$ as generic indices, 
whereas $i,j,k,l,m,...$  
are  assumed to take values on $\Sigma_{\k}^n$. A bar 
 denotes tensors and operators on $\Sigma_{\k}^n$. \\
The non-zero Riemann tensor components of the metric (\ref{b}) are: 
\begin{eqnarray} \nonumber
{R_{tr}}^{tr} &=& \frac{-2 f'' f g +f'^2g+f' g' f}{4f^2g^2} \\ \nonumber
{R_{ij}}^{kl} &=& \left( \frac{\k g-1}{r^2g} \right) 
(\d^k_i \d^l_j - \d^k_j \d^l_i)\\  \label{rie}
R_{it}{}^{jt} &=& \frac{-f'}{2rfg} \d_i^j \\  \nonumber
R_{ir}{}^{jr} &=& \frac{g'}{2rg^2} \d_i^j   \nonumber
\end{eqnarray}
The   non-zero Ricci tensor components are
\begin{eqnarray} \nonumber
R_t{}^t &=& \frac{-2 f''
 f g +f'^2g+f' g' f}{4f^2g^2} - \frac{nf'}{2rfg} \\ \nonumber
R_r{}^r &=& \frac{-2 f'' f g +f'^2g+f' g' f}{4f^2g^2} + \frac{ng'}{2rg^2}\\
R_i{}^j &=& \frac{r g' f - r f' g + 2 g f (\k g-1)(n-1)}{2 r^2 g^2 f} \d_i^j
\label{Ricbh}
\end{eqnarray}
We study  perturbations around (\ref{b}) of the form,
\begin{equation} \label{p1}
g_{ab} \to g_{ab} + h_{ab}.
\end{equation}
Indices of $h_{ab}$  are   raised using the background 
metric, therefore
$\d g^{ab} = - h^{ab}$. 
The first order variation of the Riemann  tensors is:
\begin{equation} \label{Rie}
\d {R_{ab}}^{cd} =  \frac{1}{2} \left\{ {R_{ab}}^{df} {h_f}^c - {R_{ab}}^{cf}
{h_f}^d + \left( \c_b \c^c {h_a}^d - \c_a \c^c {h_b}^d  \right) 
+ \left( \c_a \c^d {h_b}^c - \c_b \c^d {h_a}^c \right) \right\} 
\end{equation}
For transverse ($\c^a h_{ab}=0$) traceless ($g^{ab} h_{ab}=0$) perturbations (\ref{Rie}) gives
\begin{eqnarray} \label{Ric}
\d {R_a}^c &=& \frac{1}{2} \left\{ -\c^d \c_d {h_a}^c - {R_a}^f {h_f}^c + {R_f}^c 
{h_a}^f -2 {R_{ad}}^{cf} {h_f}^d \right\}\\
\d R &=&  -R_d{}^f h_f{}^d
\end{eqnarray}
from where
\begin{equation} \label{pRic}
\d R_{ab} = \d {R_a}^c g_{cb} + {R_a}^c h_{cb} = -\frac{1}{2} \c^d \c_d h_{ab} 
+ \frac{1}{2} \left( {R_a}^f h_{bf} + {R_b}^f h_{af} \right) - R_{akbf} h^{fk} 
\equiv \frac{1}{2} (\Delta_L h)_{ab}
\end{equation}
$ \Delta_L$ being the Lichnerowicz operator.
The transverse traceless condition does not restrict the perturbation,
it (partially) fixes the gauge. Linear perturbations can be classified 
into tensor, vector, and scalar 
perturbations \cite{koda}. Tensor perturbations are specific of 
higher dimensional ($D>4$) spacetimes and are the ones studied in this paper.
They satisfy $h_{a b}=0$ unless 
$(a,b)=(i,j)$.
The non-zero components $\c_a h_{bc}$ for such a tensor are
\begin{equation}\label{p2}
\c_t h_{ij} = \p_t h_{ij} \hspace{1cm} \c_r h_{ij} = \p_r h_{ij} 
- \frac{2}{r} h_{ij} \hspace{1cm} \c_i h_{jr} = -\frac{1}{r} h_{ij}
\hspace{1cm} \c_i h_{jk} = \bar \c_i h_{jk},
\end{equation}
Thus, tensor perturbations  satisfy the conditions 
$\bar g ^{ij} \bar \c_i  h_{jk} = 0 $ and $\bar g^{ij}   h_{ij} = 0$ 
(transverse traceless on $\Sigma_{\k}^n$) if and only if 
$g^{ab} \c_a  h_{bc} = 0 $ and 
$g^{ab}   h_{ab} = 0$ (transverse traceless
 on the spacetime).
Since transverse traceless tensors (TTT) on $\Sigma_{\k}^n$ can be expanded using 
a basis of eigentensors
of the Laplacian, we need only 
consider TTT perturbations of the form 
\begin{equation} \label{pert}
h_{ij}(t,r,x) = r^2 \phi(r,t) \bar h_{ij}(x) 
\end{equation}
where $r^2$ is factored for later convenience, and 
\begin{equation} \label{p3}
\bar \c^k \bar \c_k \bar h_{ij} = \g \bar h_{ij}, \;\;
 \bar \c^i  \bar h_{ij} = 0, \;\;
 \bar g ^{ij}  \bar h_{ij} = 0
\end{equation}
Note that, since $\Sigma_{\k}^n$ is a manifold of constant curvature, 
 an eigentensor of the Laplacian on $\Sigma_{\k}^n$ with 
eigenvalue $\g$ is 
also an eigentensor of $\bar \Delta_L$, 
the Lichnerowicz operator on $\Sigma_{\k}^n$ \cite{gh},  
with eigenvalue $\lambda$ given by 
\begin{equation}\label{eig}
\l = 2 \k n - \g
\end{equation}
Solutions to equation (\ref{p3}) in the case $\Sigma_{\k}^n=S^n$ 
can be obtained from \cite{higu}.
From equations (\ref{p1})-(\ref{p3})  we get 
the non trivial components of the variations of the Riemann tensor,
the Ricci tensor and the Ricci scalar. These are  displayed  in Appendix A.

\section{Tensor perturbations in EGB gravity}
\label{egb}

The Einstein-Gauss-Bonnet (EGB)  vacuum equations are
\begin{equation} \label{lovelock}
0 = {\cal{G}}_b{}^a \equiv \Lambda {G_{(0)}}_b{}^a + {G_{(1)}}_b{}^a 
+ \a {G_{(2)}}_b{}^a 
\end{equation}
Here $\Lambda$ is the cosmological constant, ${G_{(0)}}_{ab} = g_{ab}$ 
the spacetime metric, $ {G_{(1)}}_{ab} = R_{ab} -\frac{1}{2} R g_{ab}$
the Einstein tensor and  
\begin{multline} \label{g2}
{G_{(2)}}_b{}^a = R_{cb}{}^{de}  R_{de}{}^{ca}  -2 R_d{}^c R_{cb}{}^{da}
-2 R_b{}^c R_c{}^a + R  R_b{}^a 
 -\frac{1}{4} \d^a_b \left(
R_{cd}{}^{ef}R_{ef}{}^{cd} - 4 R_c{}^d R_d{}^c + R^2 \right)  
\end{multline}
the quadratic  Gauss-Bonnet tensor. These  are 
 the first  in a tower 
${G_{(s)}}_b{}^a, s=0,1,2,3,...$ of tensors  of order $s$ 
in $R_{ab}{}^{cd}$  given  by Lovelock in \cite{Lovelock}.
 As shown in  \cite{Lovelock},
the most general rank two, divergence 
free symmetric tensor that can be constructed out 
of the metric and its first two derivatives in a spacetime of dimension $d$, 
is a linear combination  of the ${G_{(s)}}_b{}^a$ with $2s <d$.
  Here we  consider 
the  static spacetimes given by  (\ref{b}). These are  
foliated 
by spacelike hypersurfaces, orthogonal to the time-like Killing vector, 
 that contain  a submanifold of dimension 
$n=D-2$ ($D$ the spacetime dimension) of constant curvature 
$\k=1,0$ or $-1$. Inserting (\ref{rie}) in (\ref{lovelock}) we find that (\ref{b}) solves the EGB equation (\ref{lovelock}) if \cite{w3}
\begin{equation}\label{f}
\frac{1}{g(r)} = f(r) = \k - r^2 \psi(r)
\end{equation}
and $\psi(r)$ satisfies 
\begin{equation}\label{p}
 \a \, P(\psi(r)) \equiv \frac{\alpha n(n-1)(n-2)}{4} \psi(r)^2 + \frac{n}{2} \psi(r) - \frac{\Lambda}{n+1} = \frac{\mu}{r^{n+1}}
\end{equation}
From (\ref{Ricbh}) and (\ref{f}), the Ricci scalar for this solution is
\begin{equation} \label{ricci}
R =  (n+2)(n+1)\psi(r)+2r(n+2) \frac{d \psi(r)}{dr}+r^2\frac{d^2 \psi(r)}{dr^2}
\end{equation}

TTT perturbations around this solution produce first order 
variations of the tensors ${G_{(s)}}_b{}^a, s=0,1,2$ which are 
trivial unless $(a,b)=(i,j)$. Setting   $g=1/f$ and using the 
equations in Appendix A gives
\begin{eqnarray}
\d {G_{(0)}}_i{}^j &=& 0 \\
\d {G_{(1)}}_i{}^j &=& \d R_i{}^j =\left[ \left( \ddot \phi - f^2 \phi'' \right)\frac{1}{2f} - \phi' \left(\frac{f'}{2} + \frac{nf}{2r} \right) + 
\frac{\phi}{2r^2} \left( 
2 \k -\g \right) \right] \bar h_i{}^j 
\end{eqnarray}
and
\begin{multline} \label{pg2}
\d {G_{(2)}}_i{}^j = \left\{ \left( \ddot \phi - f^2 \phi'' \right)  \left(\frac{n-2}{2 r^2 f} \right)  \left[ -r f' + (n-3)(\k-f)  \right]  \right. \\ 
 \left. + \phi' \left(\frac{n-2}{2 r^3} \right) \left\{ (n-3) \left[(n-2)(f-\k) f- \k r f' \right] + r^2 (f'^2 + f'' f) +
(3n-7) r  f' f 
 \right\}  \right. \\ 
\left.  + \phi \left(\frac{\g-2\k}{2 r^4} \right) \left[ r^2 f'' + 2 (n-3) r f' + (n-3)(n-4)(f-\k) \right] 
 \right\} \bar h_i{}^j 
\end{multline}
For later simplicity, we introduce three  functions 
$K_j(r)$, defined by  
\begin{equation}
\d {G_{(2)}}_i{}^j =  \left\{  \left( \ddot \phi - f^2 \phi'' \right) K_1 +
\phi' K_2 + \phi K_3 \right\} \bar h_i {}^j.
\end{equation}
Perturbations around a solution of (\ref{lovelock}) satisfy the equation
\begin{equation}
\d {G_{(1)}}_a{}^b + \a \d {G _{(2)}}_a{}^b = 0 
\end{equation}
which, after setting $\phi(r,t) = e^{\w t} \chi(r)$ gives a second order ODE 
for $\phi(r)$
\begin{eqnarray}
0 &=& -f^2 \chi''(r) + p(r) \chi'(r) + (q(r)+\w ^2) \chi(r) \\
p &\equiv& \frac{2 \a r f K_2 - r f f' - n f^2}{r+ 2  \a r f K_1}\\
q &\equiv& \frac{ 2 \a r^2f K_3 + (2 \k -\g) f}{r^2+ 2 \a r^2 f K_1} \label{q}
\end{eqnarray}

By further introducing, 
\begin{equation}
\label{factor} 
\Phi(r) = \chi(r) K(r) 
\end{equation}
with,
\begin{equation}
\label{k}
K(r) = \exp \left( - \frac{1}{2} \ln(f) - \int^r \frac{p}{2f^2}\, dr \right)
\end{equation}
and switching to ``tortoise" coordinate $r^*$, defined by ${dr^*}/dr=1/f$, 
this ODE can be cast in the Schr\"odinger form,
\begin{equation} \label{schro}
- \frac{d^2 \Phi}{{dr^*}^2}  + V(r(r^*)) \Phi = - \w ^2 \Phi \equiv E \Phi
\end{equation}
The spacetimes (\ref{b})  will therefore be stable if (\ref{schro}) has no negative eigenvalues.
On the other hand, properly normalized  eigenfunction of (\ref{schro}) with 
suitable boundary conditions (see, e.g. \cite{gh} for details) having a 
 negative eigenvalue ($E < 0)$, signals the possibility of an 
instability.
The explicit form of $K(r)$ is 
\begin{equation}
\label{k2} 
K(r) = r^{n/2-1} \sqrt{r^2+\alpha (n-2)\left((n-3)(\k -f)-r \frac{df}{dr} \right)} 
\end{equation}
The explicit form of the potential $V(r)$ as a function of $r$ and the parameters of the 
theory is rather lengthy. We notice however that the function $q$ in (\ref{q})
is,
\begin{equation} \label{v2}
q = \left(\frac{f (2 \k -\gamma)}{r^2} \right) \left(  \frac{(1-\alpha f'')r^2 +\alpha (n-3) \;
 [(n-4) (\k-f)-2r  f' ]}{r^2 +\alpha (n-2) \left[ (n-3)(\k -f)-r f' \right]}
\right)
\end{equation}
and the potential is given by,
\begin{equation}\label{v1}
V(r) = q +   \frac{f}{K} \, \frac{d}{dr} \left(f \frac{dK}{dr}  \right)
\end{equation}
V(r), given by (\ref{v1}) is the exact potential of the Schr\"odinger-like 
stability equation for the spacetime (\ref{b}) in 
EGB gravity. This includes EGB blackholes of arbitrary mass and 
cosmological constant, as well as  cosmological 
solutions of the EGB equations that result by setting $\mu=0$ in (\ref{p}). 
It generalizes the $\k=1$ case first presented in \cite{dg}, and 
it is
readily seen to reproduce the potentials  in \cite{gh} in the $\alpha=0$ (Einstein gravity) limit, a case that was 
extensively studied by Kodama and Kodama and Ishibashi (see, e.g., \cite{ki} 
and references therein). The restricted cases in \cite{n} and \cite{bd} 
can also be studied using (\ref{v1}).

\section{Classification of maximally symmetric static solutions}
\label{cms}

A classification scheme for the solutions of the EGB equations 
is introduced below following Whitt \cite{w3}. It should be 
kept in mind that a particular EGB theory is defined once the values 
of the space-time dimension $n+2$, the cosmological constant $\L$ 
 and $\a$ (assumed different from zero),
are given. A particular
symmetric solution (\ref{b})-(\ref{f})-(\ref{p}) of an EGB   theory further requires the specification 
of the discrete index $\k$ and of the integration constant $\mu$ in (\ref{p}).
Solutions are  classified according to their singularities, 
horizons and asymptotic behaviors. 
To analyze singularities we rewrite 
the Ricci scalar (\ref{ricci}) entirely in terms of $\psi$. 
This is done
using (\ref{p}) and its first two derivatives together with (\ref{ricci}). 
We arrive at 
\begin{multline} \label{rp}
R = \left\{ n(n-1)(n-2) \; \left[ \,n^2(n+3)(n+1)(n-1)^2(n-2)^2 \a^2 \psi^4 
\right. \right. \\ \left. \left.
+ 4 n^2(n+1)(n-1)(n-2) \a \psi^3
+ 8n(3+2n)(n-1)(n-2) \L \a \psi^2 -
2n^3(n+1) \psi^2 
\right. \right. \\ \left. \left.
+ 16n (3+2n) \L  \psi 
-16 \Lambda^2  \right] + 8n(n+2) \L \right\}/ \left\{ 32 P'(\psi)^3 \right\}
\end{multline}
This form of the Ricci scalar shows that the singular points $r_{sing}$ 
of a given solution 
(\ref{f})-(\ref{p}) of the EGB equations either satisfy 
$\lim_{r \to r_{sing}} \psi(r) = \pm \infty$ or 
$\lim_{r \to r_{sing}} \psi(r) = \psi_o$, $\psi_o$ being the 
stationary point  of 
$P(\psi)$. If $\m=0$ then $\psi(r)=constant$ 
and the horizon is trivially found. In the 
$\k=0, \mu \neq 0$ case there will be a horizon only if $\psi=0$, 
which requires that 
$\m$ and $\L$ have opposite signs. The horizon will be at 
\begin{equation}
r_h = (-(n+1)\mu/\L)^{1/(n+1)} \hspace{1cm} (\k = 0, \mu \neq 0)
\end{equation}
If $\m \neq 0, \k = \pm 1$, there is a horizon at every point where 
\begin{equation}
sgn(\psi) = \k  \;\; \text{ and } 
P(\psi) = \frac{\mu}{\a} | \psi | ^{\frac{n+1}{2}} \;\; (\k=\pm 1, \m \neq 0), \label{hh}
\end{equation}
For later convenience, we re-write  (\ref{p}) as
\begin{equation} \label{pp}
 P(\psi) = \frac{n(n-1)(n-2)}{4} (\psi-\L_1)(\psi-\L_2)
= \frac{\m}{\a r^{n+1}}
\end{equation}
where

\begin{equation}
\L_i = \frac{1}{\a (n-1)(n-2)} \left( -1 \pm \sqrt{1 + \frac{4 \a \L (n-1)
(n-2)}{n(n+1)}} \;\; \right)
\end{equation}
Note that, for $\m/\a > 0$ the condition  $f=\k-r^2 \psi > 0$ reduces to
\begin{eqnarray} \label{c1}
\psi &\leq & 0 \;\; \text{ or } \;\; 0< \psi, \; \frac{\m}{\a} 
|\psi|^{\frac{n+1}{2}} \leq P(\psi) \;\; \text{ ( if } 
 \;\; \k=1 ) \\ \label{c0}
\psi &\leq&  0  \;\; \text{ ( if } \;\; \k=0 )\\
\psi &\leq& 0 \;\; \text{ and  } \;\;\frac{\m}{\a} |\psi|^{\frac{n+1}{2}}
 \geq P(\psi) \;\; \text{ ( if } 
 \;\; \k=-1 ) \label{c-1}
\end{eqnarray}
whereas for   $\m/\a < 0$, $f=\k-r^2 \psi > 0$ is
equivalent to 
\begin{eqnarray} \label{cc1}
\psi &\leq & 0 \;\; \text{ or } \;\; 0< \psi, \; P(\psi) \leq \frac{\m}{\a} 
|\psi|^{\frac{n+1}{2}} \;\; \text{ ( if } 
 \;\; \k=1 ) \\ \label{cc0}
\psi &\leq&  0  \;\; \text{ ( if } \;\; \k=0 )\\
\psi &\leq& 0 \;\; \text{ and  } \;\;P(\psi) 
\geq \frac{\m}{\a} |\psi|^{\frac{n+1}{2}} \;\; \text{ ( if } 
 \;\; \k=-1 ) \label{cc-1}
\end{eqnarray}

\noindent
We label solutions  with a three digit number in the form $a.b.c$, with $a,b$ 
and $c$ labeling the distinct ranges of values for $\a$, $\L$ 
and $\m$ respectively.
 Only the $1.1.c$ cases (positive $\a$ and $\L$) 
will be analyzed in full detail, 
since the other cases are trivial variations of this one. 
A plot of  $P$ and  $\mu |\psi|^{(n+1)/2}/\a$ vs $\psi$ is 
given in each case; the positive (negative)
 $\psi$ 
intersections of these curves 
are $\k=1$ ($\k=-1$) horizons, $\psi=0$ being the horizon 
when $\k=0$. If $\m/\a > 0$ ($<0$), the portion of $P$ above (below) 
the $\psi$ axis gives the two solution 
branches $\psi_i(r)$ of (\ref{p}). Note from (\ref{p}) that  $r$ can 
extends to infinity only if $P$ has real roots, and that there is a 
singularity at $r_{sing}$ if either $\lim_{r \to r_{sing}} \psi(r) = \pm
\infty$ or $\lim_{r \to r_{sing}} \psi(r) = \psi_o$, $P'(\psi_o)=0$. 
Equations (\ref{c1})-(\ref{c-1}) ((\ref{cc1})-(\ref{cc-1})) are  used
to find the $f \geq 0$ region of interest when $\m/a > 0$ ($<0$).\\

\noindent
{\bf Case 1: $\a > 0$}\\
{\bf Case 1.1: $\L > 0$} \\
In this case $P$ has two real roots $\L_1 < 0 \leq \L_2$, $|\L_2| < |\L_1|$.
If $\m>0$ (\ref{pp}) has two solutions $\psi_i(r)$, $i=1,2,$ 
 with 
$r$ extending  to infinity, and $\lim_{r \to \infty} \psi_i(r)=\L_i$
(Figure 1.a). For $\psi_1(r)$, 
 as $r$ goes from infinity down to 
zero, $\psi$ runs from $\L_1$ to $-\infty$, where, according to our previous analysis, there is a curvature singularity. Similarly, for 
 $\psi_2(r)$, as $r$ goes down to 
zero,  $\psi$ runs from $\L_2$ to $+\infty$, where there is a curvature
 singularity. Three qualitatively different  $\m>0$ cases are 
plotted in Figure 1.b. The positive $\psi$ intersections with $P$ give 
horizons in the $\k=1$ case (eq.(\ref{hh})), the negative
 $\psi$ intersections with $P$ give 
horizons in the $\k=-1$ case, and $\psi=0$ is the $\k=0$ horizon. Note 
from (\ref{hh}) that 
some of the drawn $\k=\pm 1$ horizons  (curve intersections) 
may be missing in the special case $n=3$. 
Note also that, as $\L \to 0, \L_2 \to 0$ and some horizons move to 
infinity.\\
\noindent
{\bf Case 1.1.i: large positive $\m$} (figure 1.a and  curve (i) of figure 1.b) \\
In view of eqns (\ref{c1})-(\ref{c-1}) the $\psi_2$ branch never gives $f>0$,
whereas, for any $\k$, the $\psi_1$ branch gives a spacetime 
with $r_{sing}=0 < r < \infty$ (naked singularity).\\
{\bf Case 1.1.ii: intermediate positive $\m$} (figure 1.a and  curve (ii) 
of figure 1.b) \\
The analysis for the $\psi_1$ branch is as in case 1.1.i. The $\psi_2$ 
branch gives $f>0$ for $\k=1$, case in which the 
spacetime has two horizons and no singularities, $r_{hor_1} < r < r_{hor_2}$.
As $\L \to 0^+$, $r_{hor_2} \to \infty$.
If $n=3$ one of the intersections of $P$ with curve (ii) may be absent, and 
$r_{sing}=0 < r < r_{hor}$. \\
{\bf Case 1.1.iii: small positive $\m$} (figure 1.a and  curve (iii) 
of figure 1.b) \\ 
The analysis for the $\psi_2$ branch is as in case 1.1.ii.
For $\psi_1$ and $\k=0,1$, $r_{sing}=0 < r < \infty$ (naked singularity), 
whereas for $\k=-1$ there are two $ f > 0$ regions, 
one for which $r_{sing}=0 < r < r_{hor_1}$ (naked singularity), the 
other satisfying    $r_{hor_2} < r < \infty$ . The first 
region may be missing if  $n=3$. \\
\noindent
{\bf Case 1.1.iv: $\m=0$} \\
We obtain cosmological, non-singular 
solutions $f(r) = 1-r^2 \L_2$, with $ 0 < r < \L_2 ^{-1/2}$, 
$f(r) = \k -r^2 \L_1$, $\k=0,1$ and $ 0 < r$,  and $f(r) = -1 -r^2 \L_1$,
($\k= -1$),
$r > |\L_1|^{-1/2}$.\\

\begin{figure}[h]
\includegraphics[width=6in]{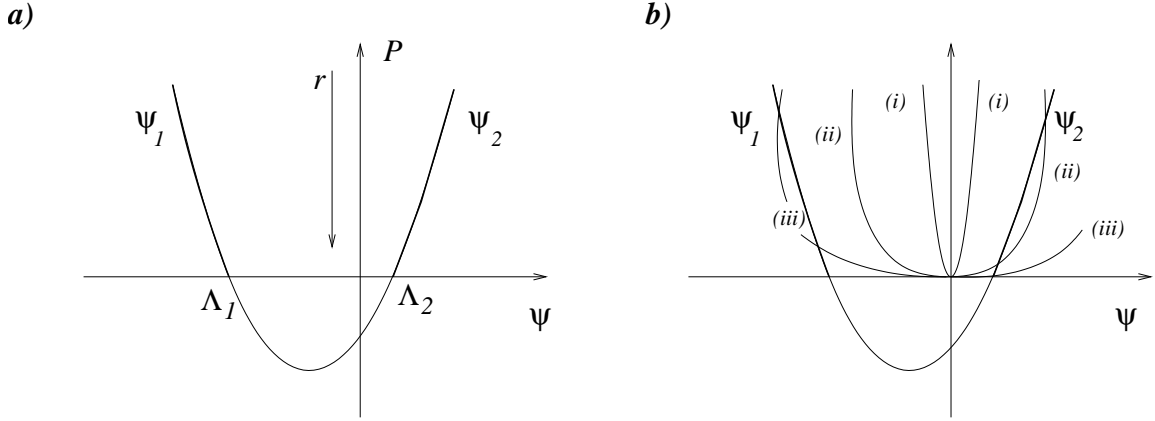}
\caption{\label{fig1} Cases 1.1.i to 1.1.iii: 
a) the two branches $\psi_i(r), i=1,2$ of equation
(\ref{pp}) in the case $\m / \a > 0$. $\psi_i \to \L_i$ as $r \to \infty$, $\psi_1$ 
($\psi_2$)  
tends to $ -\infty$ ($+\infty$)  as $r \to 0^+$. b) Plots of 
$\m |\psi|^{(n+1)/2}/\a$ for (i) large (ii) intermediate and (iii) small
positive $\m/\a$.
}
\end{figure}

If $\m<0$ (\ref{pp}) has two solutions $\psi_i(r)$, $i=1,2,$ 
 with 
$r$ extending  to infinity, and $\lim_{r \to \infty} \psi_i(r)=\L_i$.
 There is a minimum value  $r=r_{sing}$ 
defined by 
$\psi _1 (r_{sing}) =  \psi _2 (r_{sing}) = \psi _o  $ ($\psi_{o} 
\equiv  (\L_1+\L_2)/2$), this point is singular 
in view of eq. (\ref{rp}) because $P'(\psi_o)=0$.
 As $r$ grows  from $r_{sing}$ to infinity, 
$\psi_i$ goes from $\psi_o$ to $\L_i$
(figure 2.a). \\
\noindent
{\bf Case 1.1.v: small negative $\m$} \\
The $\psi_1$ branch has, for $\k=-1$, a horizon that hides a singularity, $f>0$
 if  $(r_{sing} < ) r_{hor} < r < \infty$, whereas  for $\k=0,1$ there is 
  a naked 
singularity, $r_{sing} < r < \infty$. The $\psi_2$ branch gives no 
$f>0$  solution 
for $\k=-1$, whereas for $\k=0,1$ gives a spacetime with $r_{sing} 
< r < r_{hor}$.

\noindent
{\bf Case 1.1.vi: large negative $\m$} \\
For any $\k$, the  $\psi_1$ branch gives a space-time 
with  a naked singularity, 
$f > 0$ for    $r_{sing} <  r < \infty$. For any $\k$, 
there is a horizon in the   $\psi_2$  branch, and $f>0$ 
for $r_{sing} < r < r_{hor}$. 

\begin{figure}[h]
\includegraphics[width=6in]{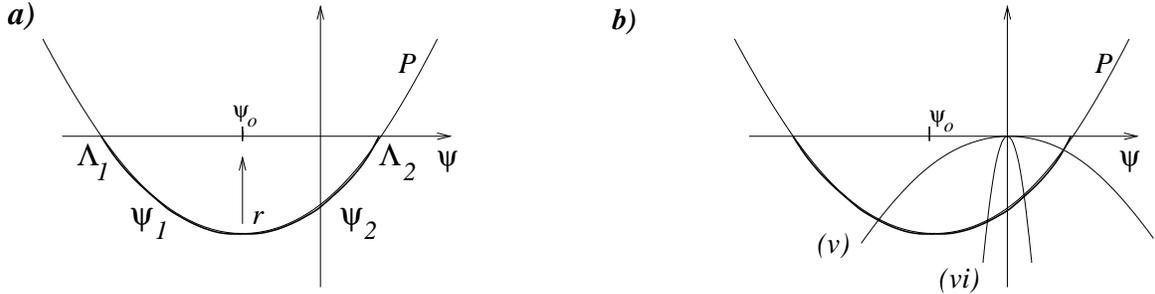}
\caption{\label{fig2} Cases 1.1.v and 1.1.vi: 
a) the two branches $\psi_i(r), i=1,2$ of equation
(\ref{pp}) in the case $\m / \a < 0$,  $\psi_i$ goes from $\psi_o$ to $\L_i$ 
as $r$ goes from $r_{sing}$ to $\infty$. b) Plots of $P$ and 
$\m |\psi|^{(n+1)/2}/\a$ vs. $\psi$ for (v) small and  (vi) large 
negative $\m/\a$.
}
\end{figure}

\noindent
{\bf Case 1.2: $-n(n+1)/ \left( 4 \a (n-1) (n-2) \right) < \L < 0$} \\
In this case $P$ has two real roots $\L_1  < \L_2 < 0$.
Six cases  of $\m$ values should be distinguished: 1.2.i large 
positive, 1.2.ii small positive, 1.2.iii null, 1.2.iv small negative, 
1.2.v intermediate negative and 1.2.vi large negative. These 
different cases are reepresented in figure 3 below.

\begin{figure}[h]
\includegraphics[width=6in]{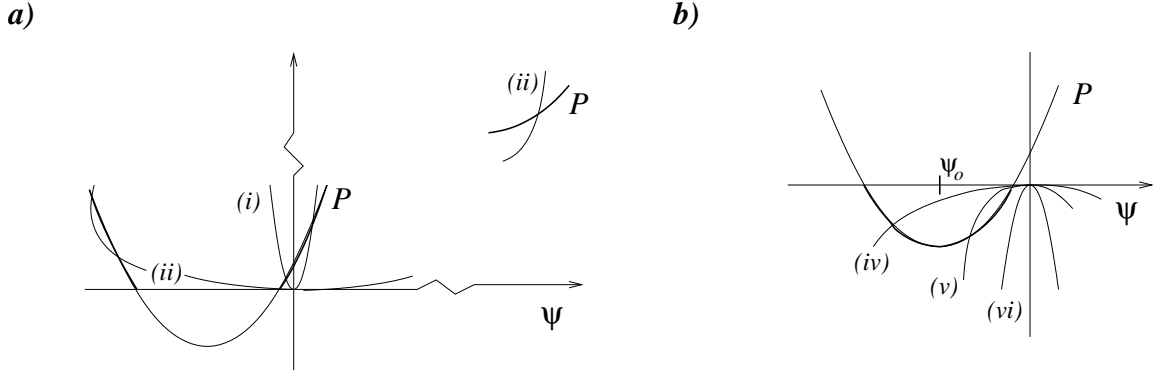}
\caption{\label{fig3} Cases 1.2.i to 1.2.vi:
 a) the two branches $\psi_i(r), i=1,2$ of equation
(\ref{pp}) in the case $\m / \a > 0$ together with the 
$\m |\psi|^{(n+1)/2}/\a$ curve for (i) large and  (ii) small
positive values of $\m/\a$.
 b) $\psi_i(r), i=1,2$ in the case $\m / \a < 0$ together with the 
$\m |\psi|^{(n+1)/2}/\a$ curve for
(iv) small  (v) intermediate and (vi) large 
negative values of $\m/\a$.
}
\end{figure}

\noindent
{\bf Case 1.3: $\L < -n(n+1)/ \left( 4 \a (n-1) (n-2) \right)$} \\
$P$ has complex roots, $\m/\a$ must be positive, and there is a maximum value
of $r$ (corresponding to $\psi=\psi_o$) which is singular.
Three ranges   of $\m$ values should be distinguished: 1.3.i large 
positive, 1.3.ii intermediate positive and 1.3.iii small positive.
These are illustrated in figure~4.

\begin{figure}[h]
\includegraphics[width=4in]{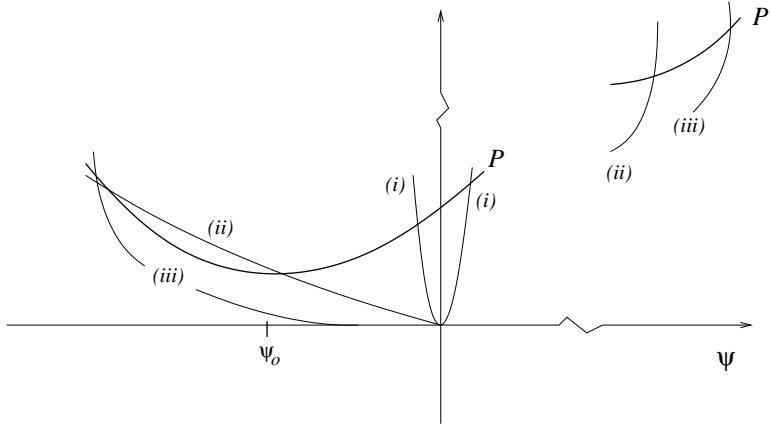}
\caption{\label{fig4} Cases 1.3.i to 1.3.iii: $P$ and $\m |\psi|^{(n+1)/2}/\a$
for (i) large, (ii) intermediate and (iii) small positive values 
of $\m/\a$. 
 $P$ has no real roots, 
as $r$ grows from $r_{sing}$,  $\psi_1(r)$ ($\psi_2(r)$) moves to 
the left (right) of $\psi_o$.
}
\end{figure}
 
\section{Stability of maximally symmetric static solutions}
\label{sms}

The stability of the solutions (\ref{b})-(\ref{lovelock})-({\ref{p}) 
of the EGB vacuum equation
can be analyzed 
using  the ``S-deformation" approach \cite{ki}: consider the operator 
\begin{equation}
A:= -\frac{d^2}{{dr^*}^2} + V
\end{equation}
acting on smooth functions defined on $I = \{ r^* |r_1^* < r^* < r^*_2 \}$, 
the regular, $f > 0$ region (note that it is possible that $r_i^* = \pm \infty$).
$E$ in (\ref{schro}) is greater than or equal to the lower bound  of 
$(\phi,A \phi)/(\phi,\phi)$, $\phi$ smooth of compact support on $I$.
However, for any such $\phi$, given a smooth $S$, 
\begin{equation} \label{sd}
(\phi,A \phi) = \int_{r_1^*}^{r^*_2}  \left( |D \phi|^2 
+ \tilde V |\phi|^2 \right) dr^*, 
\end{equation}
where
\begin{equation}
D = \frac{d}{dr^*} + S, 
\end{equation}
and the ``deformed potential" $\tilde V$ is 
\begin{equation} \label{vt}
 \tilde V = V + f \frac{dS}{dr}-S^2
\end{equation}
If an $S$ function is found such that $\tilde V \geq 0$ on $I$, the stability 
of the solution is guaranteed, as follows from (\ref{sd}).
Note from (\ref{v1}) that the choice 
\begin{equation} \label{s}
S = - f \frac{d}{dr} \ln(K)
\end{equation}
gives  $\tilde V = q$, then
\begin{equation}
(\phi,A \phi) = \int_{r^*_1}^{r^*_2}   |D \phi|^2 dr^*
+ \int_{r_1}^{r_2} \frac{|\phi|^2 q}{f} \, dr.
\end{equation}
Defining 
\begin{equation}\label{H}
H \equiv \frac{r^2 q}{f (2\k-\g)}
\end{equation}
the expectation value of $A$ can be conveniently written as
\begin{equation} \label{exA}
(\phi,A \phi) = \int_{r_1^*}^{r^*_2}   |D \phi|^2 dr^*
+ (2\k-\g) \int_{r_1}^{r_2} \frac{|\phi|^2  H}{r^2} \, dr, 
\end{equation}
Note that neither $H$ nor $D$ depend on $\g$. This factorization of 
the ``deformed potential" $q$ is the one referred to in Section \ref{intro}, 
and is crucial to arrive at the stability criterion below. \\
If the Riemannian manifold $\Sigma^n_{\k}$ is compact without  boundary, 
applying Stokes's theorem to 
\begin{equation}
0 \leq \int_{\Sigma^n_{\k}} \left(\bar \c^i h^{jk}  - \k \bar \c^j h^{ik} \right) 
\left(\bar  \c_i h_{jk}  - \k \bar \c_j h_{ik} \right)
\end{equation}
and using the TT condition of $h_{ij}$ together with  $\bar R_{ijkl} = \k (\bar g _{ik}
\bar g _{jl} - \bar g _{jk}\bar g _{il} )$ we arrive at 
\begin{equation}
\g \leq -\k^2 n \;\; \Rightarrow \;\;  2\k-\g \geq \k^2 n + 2 \k \geq 0
\end{equation}
for $n \geq 3$. Then from (\ref{exA}) we conclude that $H \geq 0$ on $I$ 
implies stability. Now suppose $H <0$ at some point in $I$, then a test $\phi$
 can be found such that 
\begin{equation} \label{exA2}
\int_{r_1}^{r_2} \frac{|\phi|^2  H}{r^2} \, dr < \; 0.
\end{equation}
The `kinetic" piece of (\ref{exA}) may certainly be larger than the 
absolute value of the integral in (\ref{exA2}),
 but (\ref{exA}) will be negative for 
sufficiently high harmonics. We conclude that {\em a solution is stable 
if and only if $H \geq 0$ on $I$.}\\

Using (\ref{pp}) and  its first $r$ derivatives, 
and introducing $\psi_o = 
(\Lambda_1+\Lambda_2)/2$ and $\Delta = (\Lambda_2 - \Lambda_1)/2$,
a simple expression for 
$H$  in terms of $\psi$ is found which is even in 
$x \equiv (\psi - \psi_o)/\Delta$:
\begin{equation} \label{hn}
H = {\frac { \left( n-3 \right)  \left( n-5 \right) {x}^{4}+2\,
 \left( n+1 \right)  \left( 2\,n-3 \right) {x}^{2} -
 \left( n+1 \right) ^{2}}{2 \left( n-2 \right) {x}^{2}
 \left( {x}^{2} \left( n-3 \right) + \left( n+1 \right) 
 \right) }}
\end{equation}
Also, if $\m \neq 0$,
\begin{equation} \label{inte}
\int_{r_1}^{r_2} \frac{H}{r^2} dr = \left(\frac{2}{n+1}\right)
\left \vert \frac{\a}{\m} \right \vert ^{\frac{1}{n+1}}  \left( \frac{n(n-1)(n-2) 
 \Delta^2}{4} 
\right)^{\frac{1}{n+1}} \int_{x_2}^{x_1} \frac{x  \left \vert 
x^2-1 \right \vert^{1/(n+1)}H}{(x^2- 1)} dx 
\end{equation}

An immediate consequence of the stability criterion above and (\ref{hn})  
is that the EGB cosmologies
are all stable against tensor perturbations, since, for $\m=0$, $\psi_i(r)
=\L_i$, then $x=\pm 1$ and $H=1$.\\
Note that the cases $n=3,4,5$ of (\ref{hn}) are special:
\begin{eqnarray} \label{n=3}
H_{(n=3)} &=& {\frac {3\,{x}^{2}-2}{x^2}} \\ \label{n=4}
H_{(n=4)} &=& \frac {-{x}^{4}+50 x^2-25}{
4{x}^{2} \left( {x}^{2}+5 \right) } \\ \label{n=5}
H_{(n=5)} &=& {\frac {  7\,{x}^{2}-3  }{{x}^{
2} \left( {x}^{2}+3 \right) }}
\end{eqnarray}

\subsection{Stability analysis}

When $P$ has real roots, $x$ is real in (\ref{hn})-(\ref{n=5}).
Furthermore
\begin{eqnarray}
H_{(n=3)} &>& 0 \;\;\text{ iff } |x| > \sqrt{\frac{2}{3}}\simeq 0.82   \label{n3}\\
H_{(n=4)} &>& 0 \;\;\text{ iff } 0.71 \simeq  \, \sqrt{15}-\sqrt{10}\;  < |x| <   \sqrt{15}+\sqrt{10} \simeq 7.03 \label{n4}\\
H_{(n=5)} &>& 0 \;\;\text{ iff } |x| > \sqrt{\frac{3}{7}} \simeq 0.65  \label{n5}\\
H_{(n>5)} &>& 0 \;\;\text{ iff } |x| >
\sqrt{\frac{(n-1)(2n-3)}{(n-3)(n-5)}}
\left(\sqrt{ 1+ \frac{\left( n-3 \right)  \left( n-5 \right)  \left( n+1
 \right) ^{2}}{(n-1)^2 (2n-3)^2}}- 1 \right)^{1/2} \label{nm5}
\end{eqnarray}
The r.h.s. of (\ref{nm5}) decreases to $\sqrt {-2+\sqrt {5}} 
\simeq 0.49$ 
as $n$ grows from $n=5$.\\

\noindent
{\bf Cases 1.1.i to 1.1.iii:}\\
All these solutions are stable if $n \neq 4$.
The stability follows from (\ref{n3}), (\ref{n4}) 
and (\ref{nm5})
above, which show that $H > 0$ if $|x| > 1$ (i.e., $\psi > \Lambda_2 $ or 
$\psi < \Lambda_1$) .
The case $n=4$ is special, as follows from (\ref{n4}) and was already noticed
for  $\k=1$ and $\Lambda=0$ in \cite{dg}. We now analyze the stability 
of every cosmological and black hole $n=4$ solution found in cases 1.1.i through
1.1.iii:\\
Cases 1.1.ii-iii, $\psi_2$ branch, $\k=1$: this black hole solution has 
$r_{sing} < r_{hor_1} < r < r_{hor_2}$ and will be stable as long as 
$x_{hor_1} \leq \sqrt{15} +\sqrt{10}$, i.e., for large enough $\mu$. 
Note that this is also true for $\Lambda = 0$ ($\Lambda_2=0$), the 
low $\mu$ instability being the one found in \cite{dg}.\\
Case 1.i.iii, $\psi_1$ branch, $\k=-1$: there are two $f>0$ regions, 
and one of them gives a  black hole solution, which will be stable as long as  
$x_{hor} \leq \sqrt{15} +\sqrt{10}$. Contrast this condition to that 
obtained before, $\k=-1$ black holes require {\em small} $\mu$ to be stable.\\

\noindent
{\bf Case 1.1.iv}\\
As explained above, $H=1$ in this case. These cosmological solutions are 
all stable.\\

\noindent
{\bf Cases 1.1.v to 1.1.vi}\\
From (\ref{n3})-(\ref{nm5}) the $\k=-1$, $\psi_1$ branch  black hole
in case 1.1.v will be stable for  $|\m/\a|$ small enough. \\

The analysis of the remaining cases can be readily done as in the 
previous cases and is left to the reader.

\section{Conclusions} \label{conc}

We proposed a classification scheme for 
  static solutions to the Einstein Gauss Bonnet  gravity 
of the form $\Sigma_{\k}^n \, \times \, {\mathbb R}^+$, $\Sigma_{\k}^n$ 
an $n-$manifold of constant curvature $\k$,  and studied their
linear stability under tensor mode perturbations. 
We found an explicit form of the potential of the Schr\"odinger 
like equation governing the time evolution of the perturbation, 
and studied its spectrum using the $S-$deformation approach. 
An  $S-$deformed Schr\"odinger potential  was found that conveniently factors 
out the eigenvalue of the laplacian on $\Sigma_{\k}^n$ associated
with the perturbation, allowing a definite classification of every space-time
into stable or unstable. Preliminary results indicate that this feature of tensor 
perturbation is shared by vector perturbations. The scalar case is still under 
investigation \cite{dg3}. 
Cosmological solutions and a variety  of Euclidean or dS black holes with 
positive curvature horizons are shown to be stable in 
space times of dimension $d \neq 6$ with a positive 
values of $\a$ -the Gauss Bonnet term coupling-.
In six dimensions, these black holes  are stable only if their masses are above a critical 
value. Black holes with  negative curvature horizons are found in 
any dimensions which 
are stable only if their masses are {\em below} a critical 
value (see \cite{npb}
for a thermodynamic instability of black holes in EGB gravity). \\
The stability of these space times 
 under vector and scalar perturbations is currently being studied.

\section*{Acknowledgments}

This work was supported in part by grants of the National University of
C\'ordoba and Agencia C\'ordoba Ciencia (Argentina). It was also supported in
part by grant NSF-INT-0204937 of the National Science Foundation of the US. The
authors are supported by CONICET (Argentina).

\appendix
\section{Linearization formulas} 
The first  order variation of the Riemann tensor, Ricci tensor and Ricci scalar 
under (\ref{p1})
can be obtained after a long calculation using (\ref{p1})-(\ref{p3}). 
These are:
\begin{eqnarray}
\d R_{ti}{}^{tj} &=&  \left( 
\frac{\ddot \phi}{2f} -\frac{f' \phi'}{4fg} \right) \bar h _i{}^j \\
\d R_{ti}{}^{rj} &=& \left( - \frac{\dot \phi'}{2g} + \left(
\frac{f'}{4fg} - \frac{1}{2gr} \right) \dot \phi \right)  \bar h _i{}^j \\
\d R_{ti}{}^{jk} &=& \frac{\dot \phi}{2 r^2} \left( \bar \c ^k \bar
h _i{}^j -  \bar \c ^j \bar h _i{}^k \right) \\
\d R_{ri}{}^{tj} &=& \left[ \left( \frac{1}{2fr}- \frac{f'}{4f^2} \right) 
\dot \phi + \frac{\dot\phi '}{2f} \right] \bar h_i{}^j   \\
\d R_{ri}{}^{rj} &=& \left[ \left( \frac{g'}{4g^2} - \frac{1}{rg}
\right) \phi' - \frac{ \phi''}{2g} \right] \bar h _i{}^j \\
\d R_{ri}{}^{jk} &=& \left( \bar \c ^k \bar h _i{}^j - \bar \c ^j
\bar h _i{}^k \right) \frac{\phi'}{2r^2} \\
\d R_{ij}{}^{tk} &=&  \frac{\dot \phi}{2f} \left( \bar \c_i \bar h_j{}^k 
- \bar \c_j \bar h_i{}^k \right)     \\
\d R_{ij}{}^{rk} &=&  \frac{-\phi'}{2g}  \left( \bar \c_i \bar h_j{}^k 
- \bar \c_j \bar h_i{}^k \right)        \\
\d R_{ij}{}^{kl} &=& \left[ \left( \frac{\k \phi}{2r^2} \right)
+ \frac{\phi'}{2rg} \right] \left( \d^l_i \bar h _j{}^k - \d^l_j
\bar h _i{}^k + \d^k_j \bar h _i{}^l - \d^k_i \bar h _j{}^l
\right) \nonumber \\ &&+ \frac{\phi}{2r^2} \left(  \bar \c_j \bar \c ^k \bar
h_i{}^l - \bar \c_i \bar \c ^k \bar h_j{}^l + \bar \c_i \bar \c ^l
\bar h_j{}^k - \bar \c_j \bar \c ^l \bar h_i{}^k \right)
\end{eqnarray}
the other components of $ \d R_{ab}{}^{cd}$ being zero. The nonzero components of the 
Ricci tensor then are 
\begin{eqnarray}
\d R_i{}^j &=&\left[ \frac{\ddot \phi }{2f} + \phi' \left(\frac{g'}{4g^2} - \frac{n}{2rg}
- \frac{f'}{4fg} \right) - \frac{\phi''}{2g} + \frac{\phi}{2r^2} \left( 
2 \k -\g \right) \right] \bar h_i{}^j,
\end{eqnarray}
Finally, 
\begin{equation}
\d R = 0 
\end{equation}
From these equations and  (\ref{g2}) follows (\ref{pg2}).

\end{document}